\documentclass[a4paper]{jpconf}
\usepackage{graphicx}
\usepackage{amsmath,amssymb}
\begin{document}

\title{Phenomenological study of texture zeros of neutrino mass matrix in minimal left-right symmetric  model}

\author{Happy Borgohain$^1$, Mrinal Kumar Das$^2$ and 
Debasish Borah$^{3,}$}

\address{$^{1,2}$ Department of Physics, Tezpur University, Napaam, Tezpur, Assam 784028, India}

\address{$^3$ Department of Physics, Indian Institute of Technology Guwahati, Assam 781039, India}

\ead{happy@tezu.ernet.in. mkdas@tezu.ernet.in, dborah@iitg.ac.in}

\begin{abstract}
We studied the phenomenological implications of texture zeros in the neutrino mass matrix of the minimal left-right symmetric model (LRSM). Since the possibility of maximum zeros reduces the maximum number of free parameters of the model making it more predictive, we considered only those cases with maximum possible texture zeros in light neutrino mass matrix $M_{\nu}$, Dirac neutrino mass matrix $M_D$ and heavy right-handed (RH) neutrino mass matrix $M_{RR}$. We then computed the correlations among the different light neutrino parameters and then the new physics contributions to neutrinoless double beta decay(NDBD) for the different texture zero cases. We find that for RH neutrino masses above 1 GeV, the new physics contributions to NDBD can saturate the corresponding experimental bound. 
\end{abstract}

\section{Introduction}

Inspite of significant experimental observations in the neutrino sector in the last few decades, there are several unknowns like the absolute scale of neutrino mass, Dirac CP phase ($\delta$), octant of atmospheric mixing angle and the ordering of light neutrino mass. Besides, the dynamical origin of light neutrino masses and their mixing is still a mystery. Several beyond Standard Model (BSM) frameworks has been proposed off-late to address these issues. Our choice is the so-called left-right symmetric model (LRSM) \cite{ex1, ex2, ex3} which appeals to be one of the most straightforward and natural extensions of the SM where the gauge symmetry of the SM is extended to $ \rm SU(3)_c\times SU(2)_L\times SU(2)_R\times U(1)_{B-L}$. The most attractive feature of the LRSM lies in its relation between the high and low mass scales of the theory. In the minimal LRSM, the light neutrino masses arise naturally from a combination of type I and type II seesaw.

In a theory that has a well-motivated underlying symmetry that gives rise to a very specific structure of the neutrino mass matrix, we can have very specific predictions for the unknown light neutrino parameters which can be tested at ongoing experiments. Here we consider such a possibility where an underlying symmetry can restrict the mass matrix to have zeros at certain specific locations also called as zero texture models. Texture zero has been realized using discrete  symmetry in several earlier works \cite{ex4, ex5}. Again LRSM using discrete A4 group has been studied in \cite{ex6}, however we see that using A4 in the framework of LRSM the relevant two zero textures in neutrino mass matrix could not be realized.

It has already been shown in several earlier works that in the diagonal charged lepton basis, the light neutrino mass matrix should  have at most two zeros. Among the fifteen possible two zero textures, only six were found to be allowed by neutrinos as well as cosmology data, which has been classified as A1, A2, B1, B2, B3 and B4 as in \cite{ex7}. Since in LRSM, several mass matrices play a role in generating light neutrino mass matrix due to the combination of type I and type II seesaw, the requirement of getting the allowed texture zeros in light neutrino mass matrix can also constrain the texture zeros of the Dirac neutrino mass $M_D$ and heavy neutrino mass $M_{RR}$. Here, we first find out the correlations among light neutrino parameters from the different texture zero conditions, some of which are shown in our paper, \cite{ex7} and then the new physics contribution to  neutrinoless double beta decay (NDBD). As these process is being probed at several experiments, this study points out the possibility of probing such scenarios at those experiments. Such aspects of probing LRSM can be complementary to the ongoing collider searches as mentioned earlier.
 
 The paper has been organised as follows. In section \ref{sec:levelb}, we discuss neutrino mass in the LRSM and the texture structures of the Dirac and Majorana mass matrices in LRSM. We then summarise the contributions to NDBD in LRSM in section \ref{sec:levelc}. We discuss our numerical analysis and results in section \ref{sec:leveld} and then finally conclude in section \ref{sec:levelf}.
	\section{ Neutrino mass in minimal left-right symmetric model}{\label{sec:levelb}}
	
	We mainly focused on the simplest left-right symmetric model also known as the minimal left-right symmetric model (MLRSM) which has been widely studied in literature. We will consider all the particle contents, as shown in \cite{ex7}. The Dirac mass terms for the leptons comes from the Yukawa Lagrangian, which for the charged leptons and neutrinos are given by,
	
	\begin{equation}
	M_l=\frac{1}{\sqrt{2}}(k_2Y_l+k_1\tilde{Y_l}), M_D=\frac{1}{\sqrt{2}}(k_1Y_l+k_2\tilde{Y_l})
	\end{equation} 
	
The light neutrino mass after symmetry breaking is generated within a type I+II seesaw as,
	\begin{equation}
	\rm M_\nu= {M_\nu}^{I}+{M_\nu}^{II}
	\end{equation}
	\begin{equation}
	M_\nu=M_{LL}-M_D{M_{RR}}^{-1}{M_D}^T
	=\sqrt{2}v_Lf_L-\frac{v^2_{\rm SM}}{\sqrt{2}v_R}Y_l{f_R}^{-1}{Y_l}^T,
	\end{equation}

\subsection{Texture Zeros in Lepton Mass Matrices of LRSM}
A detailed classification of the texture zeros in the framework of LRSM is shown in our paper \cite{ex7}. Since the light neutrino mass comes from a combination of type I seesaw term $M_D{M_{RR}}^{-1}{M_D}^T$ and a type II seesaw term $M_{LL} \propto M_{RR}$, the requirement of having allowed number of zeros in $M_\nu$ can constrain the texture zeros in $M_D, M_{RR}$. For our  numerical analysis, we will consider 4-0 texture $M_{RR}$ and  5 zeros  texture $M_D$ which can phenomenologically provide the allowed zero textures in the light neutrino mass matrix. 

\begin{table}[h!]
		\caption{\label{ex1}Number of different textures obtained for 5-0 $M_D$, 4-0 $M_{RR}$ (with rank 3). A and NA in brackets represent allowed and not allowed cases.}
			\begin{center}
	\begin{tabular}{llllllll}
		\hline
		$M_{RR}$&1-0(A)&	2-0(A) & No-0(A)   & 2-0(NA)& 3-0(NA)&4-0(NA)&Total $M_D$\\ \hline
		1&20 &27	&6 &48  &23 &2&126\\ \hline\hline\hline
		2&20 &27  	&6 &51 &20  &2&126\\ \hline\hline\hline
		3&22 &55    &6 &21 &20 &2&126\\ \hline	
	\end{tabular}
	\end{center}	
\end{table}

\section{Neutrinoless Double Beta Decay (NDBD) in LRSM}{\label{sec:levelc}}

NDBD is a  process that violates lepton number by two units and hence is a probe of Majorana neutrinos, which are predicted by generic seesaw models of neutrino masses. NDBD can be used to discriminate between neutrino mass orderings besides probing the intrinsic nature of the neutrinos and the absolute scale of the neutrino mass. If light neutrinos are Majorana, we can get a sizable contribution to NDBD especially when the ordering is of inverted type. In LRSM, due to presence of several new heavy particles, several new physics contribution to NDBD amplitude arises. The detailed analysis has been shown in \cite{ex7}. The total analytic expression for the inverse half-life governing NDBD considering all the dominant contributions that could arise in LRSM is given by,
\begin{equation}\label{eq48}
\begin{split}
\left[{T_{\frac{1}{2}}}^{0\nu}\right]^{-1}=G^{0\nu}(Q,Z)\left({\left|M^{0\nu}_\nu\eta_\nu+M^{0\nu}_N\eta_{N_R}^L\right|}^2+{\left|M^{0\nu}_N\eta_{N_R}^R+M^{0\nu}_N\eta_{\Delta_R}\right|}^2+{\left|M^{0\nu}_\lambda\eta_\lambda+M^{0\nu}_\eta\eta_\eta\right|}^2\right),
\end{split}
\end{equation}
where, $G^{0\nu}(Q,Z)$ represents the phase space factor and $M^{0\nu}$ is the nuclear matrix element which have different values for different contributions as shown in \cite{ex7}.
\section{Numerical Analysis and Results}{\label{sec:leveld}}

\begin{figure}[h]
\begin{minipage}{11.2pc}
\includegraphics[width=11.2pc]{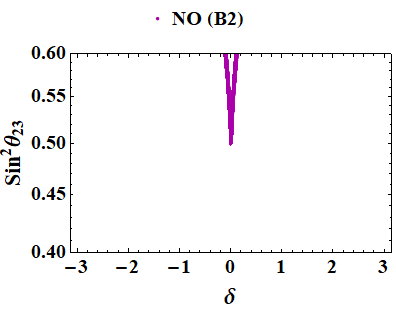}

\end{minipage}\hspace{2pc}%
\begin{minipage}{11.2pc}
\includegraphics[width=11.2pc]{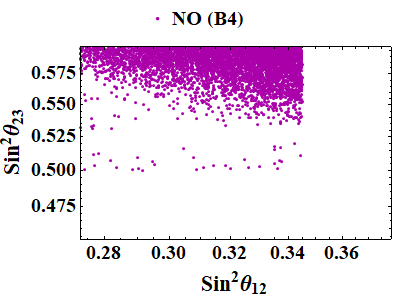}

\end{minipage} 
\begin{minipage}{11.2pc}
	\includegraphics[width=11.2pc]{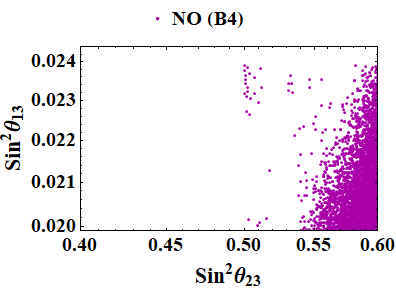}

\end{minipage} 
\caption{\label{fig1}Correlation between different neutrino parameters for NH}
\end{figure}

\begin{figure}[h]
	\begin{minipage}{11.2pc}
		\includegraphics[width=11.2pc]{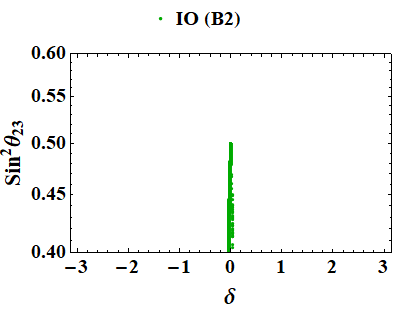}
	
	\end{minipage}\hspace{2pc}%
	\begin{minipage}{11.2pc}
		\includegraphics[width=11.2pc]{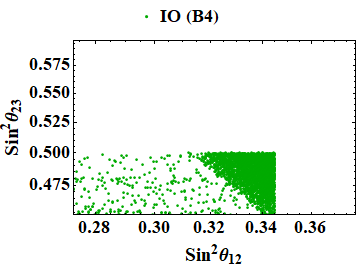}
	
	\end{minipage} 
	\begin{minipage}{11.2pc}
		\includegraphics[width=11.2pc]{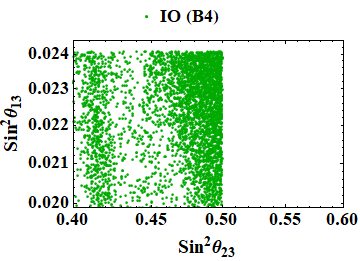}
	
	\end{minipage} 
\caption{\label{fig2}Correlation between different neutrino parameters for IH}
\end{figure}

\begin{figure}[h!]
	\centering
	\includegraphics[width=0.3\textwidth,height=3.5cm]{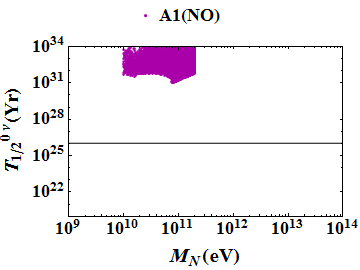}
	\includegraphics[width=0.3\textwidth,height=3.5cm]{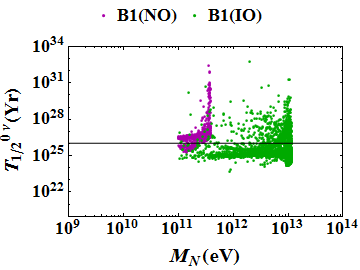}
	\includegraphics[width=0.3\textwidth,height=3.5cm]{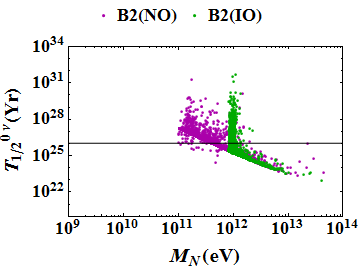}\\
	\includegraphics[width=0.3\textwidth,height=3.5cm]{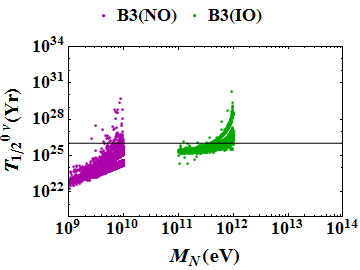}
	\includegraphics[width=0.3\textwidth,height=3.5cm]{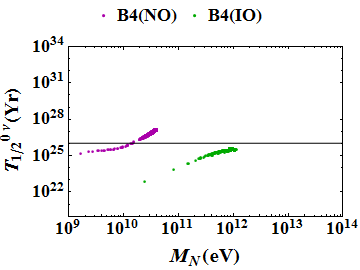}	
	\caption{Total  contribution to NDBD as a function of the lightest RH neutrino mass.} \label{fig3}
\end{figure}

In the diagonal charged lepton basis, considered in this work, we can write the light neutrino mass matrix as, $M_\nu= U_{\rm PMNS}{M_\nu}^{(\rm diag)} {U_{\rm PMNS}}^T$, 
where ${M_\nu}^{(\rm diag)} = {\rm diag}(m_1, m_2, m_3)$ and $U_{PMNS}$ is the leptonic mixing matrix. Implementing the texture zero conditions on the $M_\nu$, we can find the allowed parameter space. Out of the nine parameters of $M_\nu$, five are fixed by experimental measurements of the two mass-squared differences and three mixing angles. The remaining unknown parameters namely, $m_{\rm lightest}, \delta, \alpha, \beta$  can be predicted by the texture zero conditions. This could be done in two zero texture cases particularly, because two texture zero conditions can give rise to four real equations that can be solved simultaneously to find four unknown parameters. We vary the five known parameters randomly in the $3\sigma$ range using the recent global fit \cite{ex8} data. We see that A2 for both NO and IO and A1 (IO) are disallowed by recent data. We consider the allowed ones for our analysis for NDBD. We show some of the correlations (not shown in our earlier work \cite{ex7}) between light neutrino parameters coming out from the two zero texture conditions in figure \ref{fig1}, \ref{fig2} for both the mass ordering. Then we show the new physics contribution to half-life governing NDBD with the lightest RH neutrino mass in figure \ref{fig3}.


	\section{ Conclusion}{\label{sec:levelf}}
	We have obtained certain correlations among the light neutrino parameters from the texture zero conditions for both the mass ordering. We have seen that for the RH neutrino masses above 1 GeV, the new physics contributions to NDBD can saturate the experimental bound propounded by the KamLAND-Zen experiment. After analysing the plots shown above we can see that in the case of B1, IO can be ruled out as it is outside the experimental bound. We can say NO to be more favorable from our results. 
\subsection{\textbf{Acknowledgments}}
DB acknowledges the support from IIT Guwahati start-up grant (reference number: xPHYSUGI-ITG01152xxDB001), Early Career Research Award from SERB, DST, Government of India (reference number: ECR/2017/001873) and Associateship Programme of IUCAA, Pune. The work of MKD is supported by the DST, Government of India under the project number EMR/2017/001436.

\smallskip
\end{document}